\documentclass[reprint, 
superscriptaddress,
 amsmath,amssymb,
 aps, physrev,
]{revtex4-2}

\usepackage{graphicx}
\usepackage{dcolumn}
\usepackage{bm}
\usepackage{hyperref}

\usepackage{placeins}
\usepackage{color}
\usepackage{lipsum}
\usepackage{caption}
\usepackage{subcaption}
\usepackage{amsmath}

\begin{document}

\preprint{APS/123-QED}

\title{\textbf{Printable Nanocomposites with Superparamagnetic Maghemite ($\gamma$-Fe$_2$O$_3$) Particles for Microinductor-core Applications} 
}%

\author{M. Zambach}
\affiliation{DTU Physics, Technical University of Denmark, 2800 Kgs. Lyngby, Denmark}

\author{M. Varón}
\affiliation{DTU Physics, Technical University of Denmark, 2800 Kgs. Lyngby, Denmark}

\author{T. Veile}
\affiliation{DTU Physics, Technical University of Denmark, 2800 Kgs. Lyngby, Denmark}

\author{B. N. Sanusi}
\affiliation{DTU Electro, Technical University of Denmark, 2800 Kgs. Lyngby, Denmark}

\author{M. Knaapila}
\affiliation{DTU Physics, Technical University of Denmark, 2800 Kgs. Lyngby, Denmark}
\affiliation{Department of Physics, Norwegian University of Science and Technology, 7034 Trondheim, Norway}

\author{A. M. Jørgensen}
\affiliation{DTU Nanolab, Technical University of Denmark, 2800 Kgs. Lyngby, Denmark}

\author{L. Almásy}
\affiliation{Institute for Energy Security and Environmental Safety, HUN-REN Centre for Energy Research, 1121 Budapest, Hungary}

\author{C. Johansson}
\affiliation{RISE Research Institute of Sweden, Sensor Systems, 417 55 Göteborg, Sweden}

\author{Z. Ouyang}
\affiliation{DTU Electro, Technical University of Denmark, 2800 Kgs. Lyngby, Denmark}

\author{M. Beleggia}
\affiliation{DTU Nanolab, Technical University of Denmark, 2800 Kgs. Lyngby, Denmark}
\affiliation{Department of Physics, University of Modena and Reggio Emilia, 41125 Modena, Italy}

\author{C. Frandsen}\email{Contact author: fraca@fysik.dtu.dk}
\affiliation{DTU Physics, Technical University of Denmark, 2800 Kgs. Lyngby, Denmark}

\date{\today}

\begin{abstract}
We here present printable and castable magnetic nanocomposites containing superparamagnetic 11$\pm$3 nm $\gamma$-Fe$_2$O$_3$ particles in an insulating poly-vinyl alcohol polymer matrix. The nanocomposites feature well-dispersed particles with volume fractions between 10 and 45 \%, as confirmed by small-angle neutron scattering. The magnetic volume susceptibility is as high as 17, together with negligible hysteresis at low frequency, and constant AC-response up to the high-kHz range. Measured hysteresis curves at 100-900 kHz with up to 110 mT induced $B$-fields in the nanocomposite show that power losses depend on $B$-field squared, and frequency to the power of 1-1.3. The only loss mechanism in the nanocomposite is hysteresis losses at $>$100 kHz frequencies, where the largest particles in the 11$\pm$3 nm distribution transition from the superparamagnetic to blocked regime. To mitigate the resulting hysteresis losses (up 10$^2$-10$^5$ kW/m$^3$) a more narrow particle size distribution could be used for future materials. The presented material is eddy current-free and easily integrated into micro-fabrication protocols, as we demonstrate by fabrication of 3-turn print circuit board based inductors with cast/manual printed nanocomposite inductor cores, on which induction has been measured up to 100 MHz.
\end{abstract}

\maketitle

\section{Introduction}
Advances in miniaturisation of handheld electronics and power converters have slowed down in recent years \cite{araghchini2013a}. The reason being that magnetic components necessary for power electronics, such as inductors, cannot be miniaturized without sacrificing efficiency because inductance scales with the size of the inductor \cite{araghchini2013a, perreault2009a}. The viable options to further shrink the inductor size are therefore either employing magnetic core materials with higher susceptibility or operating the inductor at higher frequency \cite{knott2013a,van2013a}. Recent advances in semiconductor technology enable efficient switching up to tens of MHz. However, there are yet no available inductor core materials that can operate efficiently at such high frequencies while maintaining high susceptibility and high saturation magnetization \cite{araghchini2013a, perreault2009a, knott2013a,van2013a,he2023a}.

Composite materials consisting of magnetic particles in a non-conducting solid matrix are promising candidates for high frequency inductor core materials, as they potentially have lower losses at high frequency than bulk materials due to lower eddy-currents \cite{FeNi3_Article_Lu,yun2014a,yun2016a,yatsugi2019a,liu2005a,kura2012a,kura2014a,yang2018a,zambach2023b}. The caveat is that for typical composite materials containing $\mu$m sized magnetic particles, the effective susceptibility is limited by particle shape, to 3 for spheres and $\sim$10 for rods, because of demagnetization effects \cite{bj2013a,anhalt2008a}. However, it has recently been shown for materials involving smaller, single-domain particles that their susceptibility is not limited by demagnetization effects in the same way \cite{kura2012a,kura2014a,yun2016a,Zambach2025-DemagPaper}. Thus, higher effective susceptibilities and lower losses than traditional composite/powder-core materials is obtainable in composites containing insulated magnetic single-domain particles.
To this end, composites comprised of superparamagnetic (SPM) single-domain nanoparticles (NPs) \cite{kin2016a,yun2016a,kura2014a,zambach2023b} are of particular interest as they can posses higher susceptibility and lower hysteresis compared to larger, blocked single-domain particles \cite{kura2012a,FeNi3_Article_Lu,yatsugi2019a,liu2005a,kura2014a, komogortsev2021a,zambach2023b,Zambach2025-DemagPaper}.
\begin{figure*}
    \centering
    \includegraphics[width=1\linewidth]{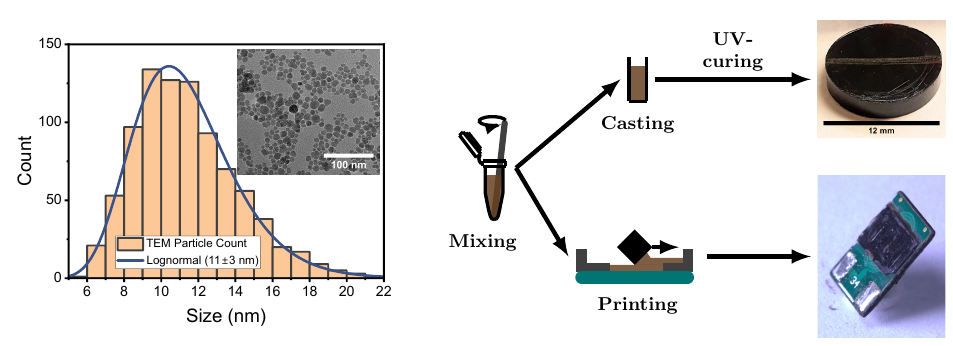}
    \caption{Nanocomposite fabrication. Solution containing superparamagnetic 11±3 nm $\gamma$-Fe$_2$O$_3$ (maghemite) particles is mixed with varying amount of poly-vinyl alcohol, see table \ref{tab:CompositeVol}, and either cast into disk shape or printed onto print circuit board (PCB) inductors.}
    \label{fig:OverviewComic}
\end{figure*}

So far, relatively low susceptibilities are reported for composites containing single-domain magnetic particles. For nanocomposites containing SPM $\gamma$-Fe$_2$O$_3$ (maghemite) particles, the highest volume susceptibilities of up to 9 are reported for dense composites \cite{yun2014a}. Theoretical predictions for materials containing SPM maghemite particles show that a susceptibility above 30 should be achievable with appropriate tuning of particle size \cite{kura2012a,zambach2023b}. Experimental proofs are, however, lacking so far due to non optimal samples, e.g. too low/high particle fraction and/or particle aggregation \cite{fabris2019a,zambach2023b}.

Ideally, the fabrication of new inductor core materials should be easily integrated into existing micro- and nano-fabrication protocols. Only very few nanocomposites developed so far offer this opportunity \cite{garnero2019a,shi2020a,hodaei2018a,yang2018a,li2022a,kin2016a,kura2012a,kura2014a,FeNi3_Article_Lu,yatsugi2019a,yun2014a,yun2016a}. Four main methods are used for fabrication of inductor cores: i) compression molding \cite{yang2018a,li2022a,kin2016a,kura2012a,FeNi3_Article_Lu,yun2014a}, ii) casting \cite{garnero2019a,kura2014a,yun2016a}, iii) screen-printing \cite{yatsugi2019a}, and iv) 3D-printing \cite{shi2020a,hodaei2018a}. While compression molding is not directly applicable for micro-fabrication, and current 3D-printing technology suffers from insufficient particle loading \cite{kin2016a,hodaei2018a,shi2020a}, casting and screen-printing of nanocomposites can be integrated directly in micro-inductor fabrication \cite{garnero2019a,kura2014a,yun2016a}. However, the materials developed for casting and printing so far have suffered from particle aggregation \cite{garnero2019a,kura2014a,FeNi3_Article_Lu} or have proven to be unstable \cite{yun2016a}. Thin film techniques are not considered here because of the low amount of material deposited.

Here, we show that it is possible to fabricate a printable nanocomposite containing SPM $\gamma$-Fe$_2$O$_3$ (maghemite) particles in a UV-cured poly-vinyl alcohol (PVA) matrix. We verify particle and nanocomposite morphology and degree of aggregation by transmission electron microscopy (TEM) and small-angle neutron scattering (SANS), and we assess the resulting susceptibility and losses by magnetic characterisation including DC-hysteresis loop measurements, AC-susceptibility measurements, high-field AC hysteresis loop measurements as well as induction measurements on fabricated printed circuit board (PCB) inductors. Power losses in the synthesised nanocomposites are measured under real working conditions with frequency ranges up to $\sim$1 MHz. We also demonstrate that the deposition/printing of the material on PCB inductors is feasible, which means easy integration into micro-fabrication protocols and allowing flexibility in the inductor design.

\section{Results}
\subsection*{Nanocomposite synthesis and morphological characterisation}
First, aqueous suspensions of well-dispersed $\gamma$-Fe$_2$O$_3$ nanoparticles were prepared by a polyol process (see details in experimental section). The particles have number-weighted lognormal distributed diameters of $11\pm3$ as determined by TEM imaging of a dried out droplet on carbon film copper grid (see example in \textbf{figure \ref{fig:OverviewComic}}). Nanoparticles in concentrated solution (44.1 mg/mL iron) were mixed with 9000-10000 M$_w$ poly-vinyl alcohol (PVA), and photo-initator Darocur 1173 was added (0.34 mL per gram PVA). Volumes are given in \textbf{table \ref{tab:CompositeVol}} in the experimental section.

pH-controlled electrostatic repulsion between the particles was used to obtain a well-dispersed suspension of the magnetic particles. A low pH of ca. 2 induced surface charges on the individual particles, resulting in a stable suspension (no precipitation of particle aggregates in years). The repulsion was kept active when mixed with the polymer, which is soluble in water.

Two types of nanocomposite samples were prepared: cast bulk samples and printed samples on PCB inductors.
For cast samples, solutions were mixed for 20 minutes, poured into mm-sized casts and placed under broad spectrum mercury UV-lamp for 2-5 hours. After UV-curing the samples were dried at moderated heat (50 \textdegree C) for two days. The cast samples are 5-12 mm in diameter and 0.2-2 mm thick. An image of cylinder/disk-shaped nanocomposite after release from the cast form can be seen in \textbf{figure \ref{fig:OverviewComic}}. The printed samples were prepared by manual printing and/or droplet casting on PCB based inductors, also seen in \textbf{figure \ref{fig:OverviewComic}}. UV-curing of printed samples mainly resulted in less brittle material. The nanocomposites have a transparent (not cloudy) red-brown colour (like glass) indicating that the NPs are not forming large aggregates ($>500$ nm) in the polymer matrix.

\begin{figure}[tp!]
    \centering
    \includegraphics[width=0.95\columnwidth]{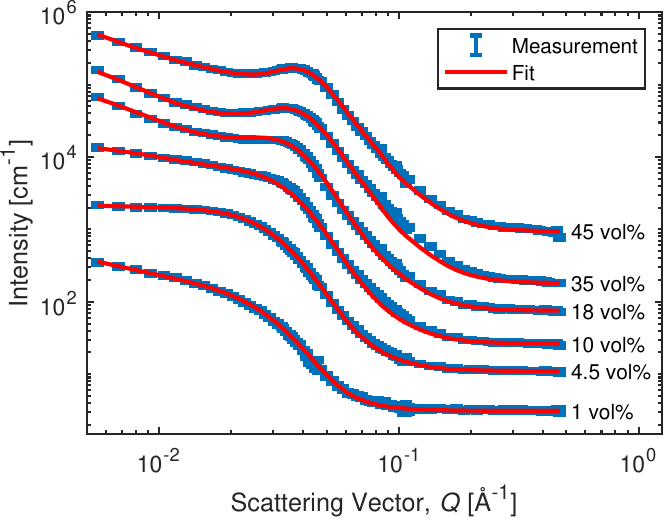}
    \caption{Reduced, background corrected and radially integrated small-angle neutron scattering data (blue) for cast samples. For better visibility, scattering intensity was scaled with $10^{0.5 x}$, $x\in[0,1,..]$ where $x$ is the sample number. Fit (red) is based on two models covering isolated and aggregated nanoparticles. Fitting models and parameters can be found in the supporting information.}
    \label{fig:SANS}
\end{figure}

Particle aggregation in the nanocomposites was investigated by SANS on cast samples \cite{alm2021a}. TEM was not possible due to charging of the non-conducting samples in the electron beam. Reduced, polymer background corrected, and radially integrated SANS data can be seen in \textbf{figure \ref{fig:SANS}}. The subtracted PVA background showed a $-4$ power law decay around 0.03-0.1 Å$^{-1}$ and an maximum at 1.4 Å$^{-1}$, and was weak compared to the nanoparticle scattering \cite{kanaya1998a}. SANS data seen in \textbf{\autoref{fig:SANS}} is thus scattering from the particles, and shows typical $-4$ power law behaviour in the 0.04-0.1 Å$^{-1}$ range with an interaction peak for high volume fractions, typical for systems containing spherical particles \cite{fournet1950a,kotlarchyk1983a}. The low $Q$ range is seen to have a relatively flat scattering intensity profile even for high volume fractions indicating good dispersion of the particles in the matrix. 

The fractions of particles in the non-aggregated and aggregated configurations were estimated by fitting the scattering data by a combination of two models. For the non-aggregated phase a spherical form-factor with interaction from a hard-sphere structure factor was used \cite{fournet1950a,kotlarchyk1983a}. For the aggregated particles, scattering was modelled by a fractal model using a spherical particle form factor and a fractal structure factor \cite{teixeira1988a}. The fits based on the two models are shown in \textbf{figure \ref{fig:SANS}}. The model functions and a full list of fitting parameters can be found in the supplementary material. 
\begin{figure}[tp!]
    \centering
    \includegraphics[width=1\linewidth]{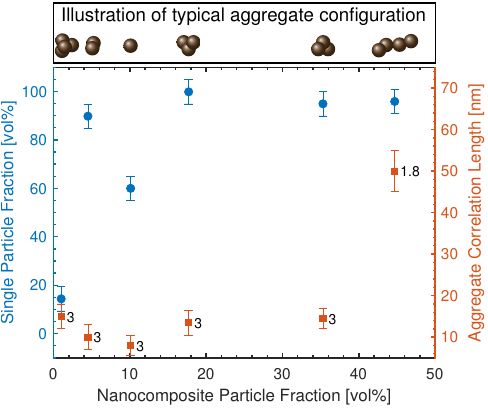}
    \caption{Single particle fraction in the nanocomposite, and correlation length of aggregates found from SANS data fitting for cast nanocomposite samples. Fractal dimension used for fit are shown as numbers beside each aggregate correlation length. Illustration of typical aggregate configuration based on aggregate correlation length and fractal dimension seen above the plot. Fitting models and parameters can be found in the supporting information.}
    \label{fig:SANS_Aggr}
\end{figure}

The two-model fit reproduce the main features of the measured samples well when using a volume weighted lognormal distributed particle diameter of around $13.8\pm2.0$ to $15.2\pm2.2$ nm, as seen from \textbf{figure \ref{fig:SANS}} and from the full model parameters given in the supporting information. Some deviation between the measured and fitted intensity is seen in the $Q$ range around 0.1-0.2 Å$^{-1}$, this may be due to insufficient polymer background subtraction or higher degree of ordering in the polymer-chains for samples with particles \cite{kanaya1998a}. 

We extract the fraction of aggregates from the relative weights of the models weight factor, see \textbf{figure \ref{fig:SANS_Aggr}}. Notably, for the 18-45 vol\% samples, $\geq$ 95\% of the NPs are non-aggregated. The smaller single particle fraction with decreasing NP volume fraction might be due to the synthesis process: in the lowest concentration composites the nanoparticle solution was diluted strongly by PVA solution, which may have increased the pH to a value insufficient to produce surface charges necessary for particle repulsion. 
\begin{table*}
\centering
 \caption{Nominal nanoparticle volume fraction, measured magnetic nanoparticle volume fraction, coercive field ($H_{\textrm{c}}$), nanocomposite volume susceptibility ($\chi_{\textrm{nc}}$), nanoparticle volume susceptibility ($\chi_{\textrm{p}}$), and Langevin fitted diameters using volume weighted, lognormal distributed mean diameter and distribution standard deviation. Results are based on room temperature DC hysteresis loops of \autoref{fig:VSM_Hyst}. For the measured nanoparticle fraction and Langevin fits, a measured particle saturation magnetisation of 303$\pm$14 kA/m is used.}
 \label{tab:VSMAnalysis}
 \begin{ruledtabular}
    \begin{tabular}{c|cccc|c}
    \textit{Nominal} & \multicolumn{4}{c|}{\textit{Demagnetisation corrected DC hysteresis measurements}} & \multicolumn{1}{c}{\textit{Langevin fitting}}\\
    \rule{0pt}{2.6ex}
    Nanoparticle & Magnetic nanoparticle & Coercivity & Nanocomposite volume & Nanoparticle volume & Lognormal mean\\
    fraction [vol\%] &  fraction [vol\%] & $H_{\textrm{c}}$ [A/m] & susceptibility $\chi_{\textrm{nc}}$ & susceptibility $\chi_{\textrm{p}}$ & diameter [nm]
    \rule[-1.2ex]{0pt}{0pt}\\
    \hline
    \rule{0pt}{2.6ex} 1.3$\pm$1 & 1.0$\pm$0.2 & 4$\pm$8 & 0.21$\pm$0.01 & 21$\pm$4 & 10.0$\pm$3.6\\
    5.7$\pm$1.7 & 4.5$\pm$0.1 & 4$\pm$8 & 1.0$\pm$0.1 & 22.2$\pm$2 & 10$\pm$3.7\\
    11.5$\pm$2.4 & 10.1$\pm$0.1 & 3$\pm$8 & 2.5$\pm$0.2 & 24.5$\pm$2.5 & 9.9$\pm$3.8\\
    19.6$\pm$2.9 & 17.7$\pm$0.5 & 3$\pm$8 & 5.4$\pm$0.5 & 30.5$\pm$3 & 10.1$\pm$4.2\\
    32.8$\pm$3.7 & 35.3$\pm$0.5 & 3$\pm$8 & 12.6$\pm$1.3 & 35.6$\pm$3.6 & 9.8$\pm$4.5\\
    39.4$\pm$5 & 44.7$\pm$0.5 & 3$\pm$8 & 17$\pm$1.5 & 38$\pm$3 & 9.6$\pm$4.2
  \end{tabular}
 \end{ruledtabular}
\end{table*}
The 10 vol\% sample shows the largest fraction of aggregates, however, this sample also shows the shortest correlation length of aggregates, allowing for 1-2 particle per aggregate. Therefore, the degree of aggregation is probably overestimated in this sample compared to the 4.5 vol\% and 18 vol\% samples. 

In all cases, the correlation length in the fractal model seems fairly stable around 10-15 nm, except for 45 vol\% where it was around 50 nm, but such increase can be explained by longer, chain-like aggregates indicated by the fractal dimension of 1.8 whereas the other samples showed fractal fitting dimension of 3, i.e. spherical aggregation (see figure \ref{fig:SANS_Aggr}). The clear deviation from the minus 3 scaling indicates that, the 45 vol\% sample does have only few more particles per aggregate than the other samples, but in a more linear configuration, resulting in a larger correlation length. 

Hence, these results reveal that the nanocomposites are well-dispersed between 20 and 45 vol\%, with above 95\% of particles being isolated and the remaining $<$5\% being in relatively small (2-5 particle) aggregates.

\subsection*{Magnetic characterisation}
\begin{figure}[bp!]
        \centering
        \includegraphics[width=1\columnwidth]{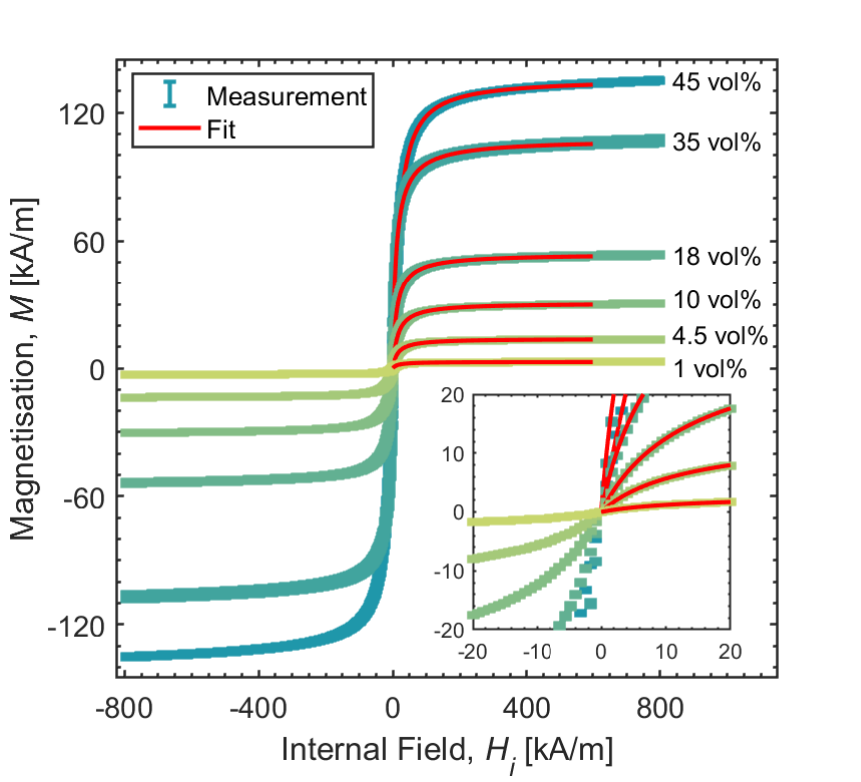}
        \caption{Room temperature DC hysteresis loop obtained by VSM for cast nanocomposite samples. Fits are based on Langevin function and volume weighted lognormal distributed particle diameters. Fitting parameters are given in table \ref{tab:VSMAnalysis}.}
        \label{fig:VSM_Hyst}
\end{figure}

\textbf{Figure \ref{fig:VSM_Hyst}} shows the DC hysteresis loops for the nanocomposites measured by vibrating sample magnetometry (VSM) and demagnetisation corrected for a flat, cylindrical sample shape according to \cite{beleggia2005a} (see \textbf{ table \ref{tab:VSMDemag}} in the experimental section). 
The magnetisation follows a typical Langevin behaviour with no coercivity or hysteresis see cf. fits in figure \ref{fig:VSM_Hyst}. The coercive field ($H_c$) at room temperature is 3-4$\pm$8 A/m for all samples as seen in the inset of figure \ref{fig:VSM_Hyst}. A summary of the magnetic characterisation and fitting parameters can be found in \textbf{table \ref{tab:VSMAnalysis}}.
The magnetic particle fraction in table \ref{tab:VSMAnalysis} was determined from VSM measurements using measured particle saturation magnetisation of 303 kA/m$^3$ (see experimental section) and the densities of maghemite and cured PVA (4.88 g/cm$^3$ and 1.00 g/cm$^3$, respectively). The measured particle fractions match well with the nominal fractions and deviations are likely due to uncertainty during pipetting.

Nanocomposite volume susceptibility is found as slope $(M,H_i) = (0,0)$ (see table \ref{tab:VSMAnalysis}) and increases stronger than linearly with volume fraction, reaching 17 for the $\sim$45 vol\% sample. The corresponding susceptibility per particle volume increases from 21 to 38 when increasing the particle fraction (see table \ref{tab:VSMAnalysis}). This is likely due to weak dipolar interactions \cite{kin2016a,elfimova2019a,chantrell2001a,usov2020a}.

Langevin fits of the DC-hysteresis curves in figure \ref{fig:VSM_Hyst} are based on volume weighted lognormal distributed particle diameters \cite{rosensweig2002a}, using the measured particle saturation magnetisation. The magnetic diameter distribution emerging from the fits are in agreement with the TEM results showing a slightly smaller mean diameter than measured by TEM (10 nm vs 11 nm), with slightly larger particle size distribution width (3.6-4.5 nm vs 3 nm; see \textbf{table \ref{tab:VSMAnalysis}}).
\begin{figure}[tp!]
    \centering
    \includegraphics[width=0.95\columnwidth]{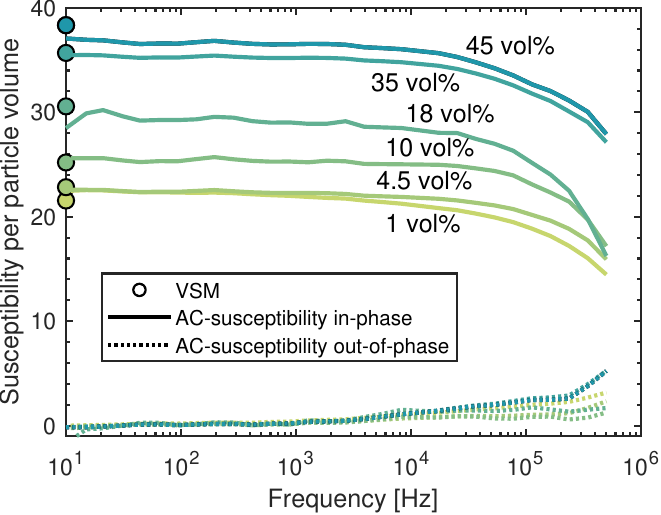}
    \caption{AC-susceptibility is shown per particle volume found by dividing demagnetisation corrected susceptibility of the nanocomposite with the volume fraction of particles. In-phase component of the susceptibility is plotted with solid lines, out-of-phase component is plotted with dashed lines, and susceptibility found from VSM-measurements are shown as circles.}
    \label{fig:AC-Susc}
\end{figure}

\textbf{Figure \ref{fig:AC-Susc}} shows the AC-susceptibility per particle volume for cast samples for frequencies between 10 Hz and 500 kHz. The per particle volume susceptibility (volume susceptibility of the nanocomposite divided by the volume fraction) increases with increasing volume fraction of particles, as seen from the DC hysteresis curves. This indicates a positive effect of a weak magnetic particle interaction, as the susceptibility per particle volume for non-interacting systems is expected to be constant \cite{zambach2023b,elfimova2019a}. 

We also observe in \textbf{Figure \ref{fig:AC-Susc}} that the in-phase component of the AC-susceptibility begins to drop around 10 kHz. This happens when the largest particles in the composite start to lose their superparamagnetism and become blocked \cite{zambach2023b}. The frequency dependence scales as the inverse of the superparamagnetic relaxation time $1/\tau_{\textrm{spm}}$, and the superparamagnetic relaxation time depends exponentially on the particle volume. Therefore, even one nm difference in diameter changes blocking frequency by orders of magnitude \cite{zambach2023b,fock2018a}. The drop in in-phase component is accompanied by an out-of-phase component increase. The frequency dependence is very similar for all samples, except the 1 vol\% and 18 vol\% that feature a slightly larger relative drop/increase in in-phase/out-of-phase component due to aggregation and lower signal-to-noise ratio. Hence, the decrease in in-phase susceptibility is mainly an effect of the particle size distribution and less due to change of superparamagnetic relaxation by inter-particle interactions. We estimate that the dipolar interaction between two 11 nm maghemite particles is on the order of 50-500 Kelvin for volume fractions between 1-45 \%, whereas the anisotropy barrier is on the order of 1200 Kelvin.

\begin{figure}[tp!]
    \centering
    \includegraphics[width=0.95\columnwidth]{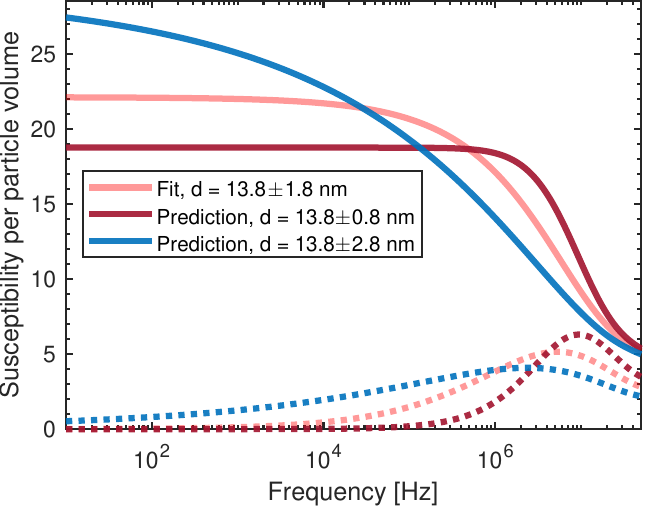}
    \caption{Debye-model predictions of the particle susceptibility for wider and more narrow particle size distributions compared to the fitting result for the 45 vol\% sample. Parameters based on fit from figure \ref{fig:SusceptibilityOverview}.}
    \label{fig:AC-Susc_Prediction}
\end{figure}

In order to optimise for more constant susceptibility and limit losses in a larger range of frequencies, it would be favourable to have more mono-disperse particles, i.e. as narrow a particle diameter distribution as possible \cite{carrey2011a,zambach2023b}. The particles should still be as large as possible while remaining SPM at the targeted applied frequency to have large susceptibility \cite{zambach2023b}.
To illustrate the effect of polydispersity on the frequency dependent susceptibility (and losses) we have made Debye-model predictions with wider and more narrow particle size distributions, see \textbf{figure \ref{fig:AC-Susc_Prediction}}. It is seen that while having a slightly lower susceptibility, a more narrow distribution will have lower losses (smaller out-of-phase component) and a more constant in-phase component. From this we expect future materials to use as monodisperse particles as possible.

\begin{figure}[tp!]
        \centering
        \begin{subfigure}{1\columnwidth}
            \centering
            \includegraphics[width=0.95\columnwidth]{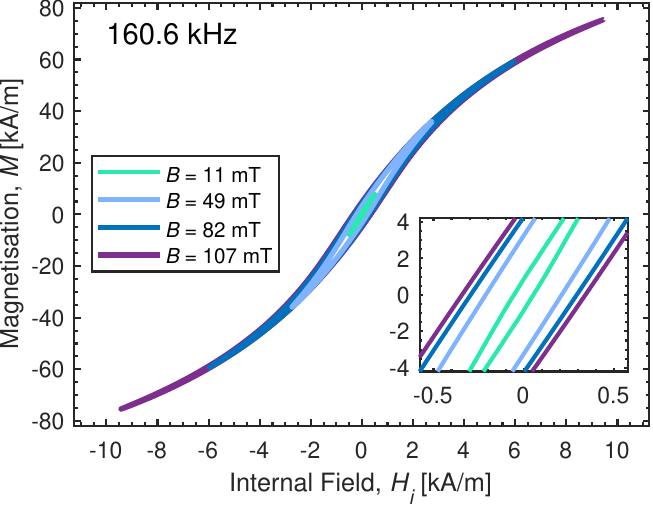}
            \caption{}
            \label{fig:LoopTracer160kHz}
        \end{subfigure}
        \vspace{0.1cm}
        
        \begin{subfigure}{1\columnwidth}
            \centering
            \includegraphics[width=0.95\columnwidth]{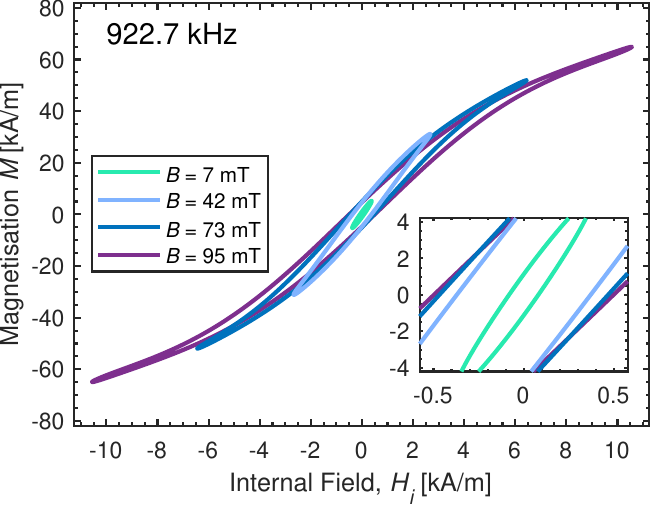}
            \caption{}
            \label{fig:LoopTracer922kHz}
        \end{subfigure}
        \caption{Demagnetisation corrected high-frequency hysteresis curves obtained at 160.6 kHz (a) and 922.7 kHz (b) for applied fields of 0.8, 4, 8, and 12 kA/m. Induced $B$-fields are 11 - 107 mT (a) and 7 - 95 mT (b).}
\end{figure}

To quantify further the losses we use high-frequency hysteresis measurements at $B$-field amplitudes relevant for inductor core materials. \textbf{Figures \ref{fig:LoopTracer160kHz}} and \textbf{\ref{fig:LoopTracer922kHz}} show hysteresis curves obtained at 160.6 and 922.7 kHz by use of an AC magnetometer (so-called looptracer) \cite{TVeile2025} for applied field amplitudes $H$ of 0.8, 4, 8, and 12 kA/m. Hysteresis curves at 241.0, 404.0, and 570.6 kHz can be found in the supplementary material. All hysteresis loops were demagnetisation corrected for sample shape \cite{Zambach2025-DemagPaper}, see table \ref{tab:AC_Demag}. For small applied field, corresponding to an induced $B$-field amplitude of 7-11 mT in the samples, the hysteresis curves are elliptical, while for larger applied fields hysteresis curves become more and more S-shaped. Accordingly, the coercivity increases for increasing fields, but is seen to somewhat saturate. Increasing the frequency increases coercivity, such that the maximum coercivity is at 350 A/m and 500 A/m for 160.6 kHz and 922.7 kHz. The slope of the hysteresis curves, i.e. the in-phase component of the susceptibility, decreases with increasing frequency and with increasing applied fields due to heat up of the sample. The reached $B$-field amplitude is therefore smaller for higher frequencies. Superparamagnetic susceptibility goes as $1/T$ and thus for higher frequencies and fields the increasing heat generated will decrease susceptibility \cite{zambach2023b}.
\begin{figure}[!bp]
    \centering
    \includegraphics[width=0.95\columnwidth]{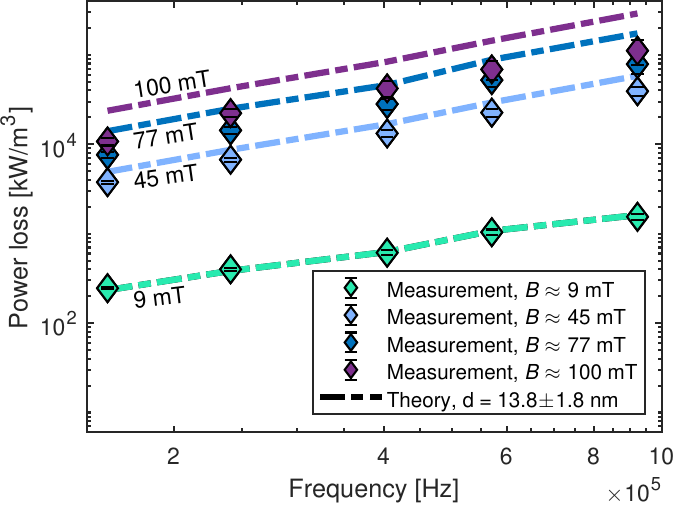}
    \caption{Power loss as function of frequency based on hysteresis area measured by the looptracer setup together with theoretic Debye-model predictions found from fitting in figure \ref{fig:SusceptibilityOverview}. Applied field amplitude is 0.8, 4, 8, and 12 kA/m, inducing roughly 9, 45, 77 and 100 mT $B$-field amplitude. Debye-model uses particle anisotropy of 12.7 kJ/m$^3$ and superparamagnetic attempt time of 2 ns as found from fitting of susceptibility in \autoref{fig:SusceptibilityOverview}, and is calculated to correspond to the $B$-field of each measurement point.}
    \label{fig:Power}
\end{figure}

From the hysteresis area measured by AC magnetometry the power losses of the magnetic material versus frequency is calculated to be in the range of $10^2$-$10^5$ kW/m$^3$, see \textbf{figure \ref{fig:Power}}. The power losses of the nanocomposite material depends on frequency with an exponent between 1 and 1.3. This is a significantly lower frequency dependence than ferrites, where eddy currents give rise to power losses with a power of 2 \cite{sanusi2022a}. Power loss measurements are shown together with predictions based on Debye-model calculations using in- and out-of-phase components of the susceptibility, where the power loss per particle is $P_{\text{p,theory}} = \mu_0 \pi f \chi''(f) H_0^2$ \cite{svedlindh1997a,zambach2023b}. Here $\chi''(f)$ is the frequency dependent out-of-phase component of the susceptibility and $H_0$ is the applied field needed to reach the induced $B$-field of the AC hysteresis curves. Predicted losses assume linear relationship between applied field and magnetisation. Parameters used in the predictions are found from fitting AC-susceptibility and AC-magnetometry data (at low fields $\leq0.4$  kA/m), as seen in \textbf{figure \ref{fig:SusceptibilityOverview}} where particle diameter of 13.8$\pm$1.8 nm, effective uniaxial particle anisotropy constant of 12.7 kJ/m$^3$, and superparamagnetic attempt time of 2 ns were found. These values are typical value for maghemite nanoparticles in the given size range \cite{carrey2011a}.

The theoretic core loss calculations of \textbf{figure \ref{fig:Power}} match the measured values within the uncertainty for low fields. A slightly larger particle size is found when compared to TEM and Langevin fits (11 and 10 nm vs. 13.8 nm), matching well with the SANS results (13.8-15.2 nm). For the largest $B$-field amplitudes the power losses are lower than the corresponding theoretic prediction due to saturation effects \cite{carrey2011a}. Heat up of the sample could also contribute to the trend that measured power-losses are below the theoretic predictions, as susceptibility and hysteresis decreases with increasing temperature \cite{zambach2023b}.

While the power losses measured are much higher compared to typical ferrite materials (roughly $10^4$ kW/m$^3$ of the here presented material at 1 MHz field and 30 mT amplitude vs 50 kW/m$^3$ for TDK's PC200 \cite{TDK,sanusi2022a}), we emphasise that the power-loss of the here presented nanocomposite does not scale with frequency squared. The almost linear frequency dependence indicates that the material has no relevant eddy current losses, as these scale with frequency squared \cite{he2023a}.
The decrease/increase of in-/out-of-phase susceptibility for our material influences the effective frequency dependence. Calculating the worst case, where susceptibility drops and thus an accordingly higher applied field is needed to obtain the similar $B$-field the exponent of frequency dependence is still below 1.5.  The main reason for the exponent of the frequency being larger than 1 is that the larger particles in the samples become blocked with increasing frequency. This however also stresses that a more narrow distribution of particle diameter would be more optimal, resulting in less decrease/increase of in-/out-of-phase susceptibility\cite{rosensweig2002a,carrey2011a,zambach2023b}, as seen in figure \ref{fig:AC-Susc_Prediction}.

\subsection*{Proof of concept: Nanocomposite containing SPM particles as printable inductor core material}
\begin{figure}[bp!]
    \centering
    \begin{subfigure}{0.49\columnwidth}
        \centering
        \includegraphics[width=1\linewidth]{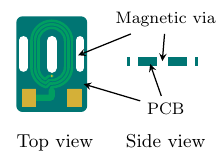}
        \caption{Without core material}
    \end{subfigure}
    \hfill
    \begin{subfigure}{0.49\columnwidth}
        \centering
        \includegraphics[width=1\linewidth]{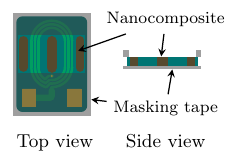}
        \caption{First deposition}
    \end{subfigure}

    \vspace{0.3cm}
    
    \begin{subfigure}{0.49\columnwidth}
        \centering
        \includegraphics[width=1\linewidth]{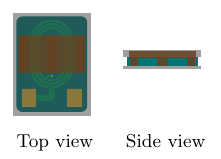}
        \caption{Second deposition}
    \end{subfigure}
    \hfill
    \begin{subfigure}{0.49\columnwidth}
        \centering
        \includegraphics[width=1\linewidth]{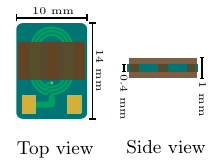}
        \caption{Finished inductor}
    \end{subfigure}
    
    \caption{Illustration of nanocomposite core deposition on print circuit board (PCB) based 3 turn inductor, shown before core deposition (a), after filling the magnetic vias (b), after top core deposition (c), and after finished core deposition (d). The finished inductor can be seen in figure \ref{fig:OverviewComic} and \ref{fig:ab}.}
    \label{fig:Printing}
\end{figure}
\begin{figure}[!tp]
\centering
\begin{subfigure}{0.79\columnwidth}
    \centering
    \includegraphics[angle=0,width=0.95\columnwidth,trim={0cm 11.5cm 2.5cm 9cm},clip]{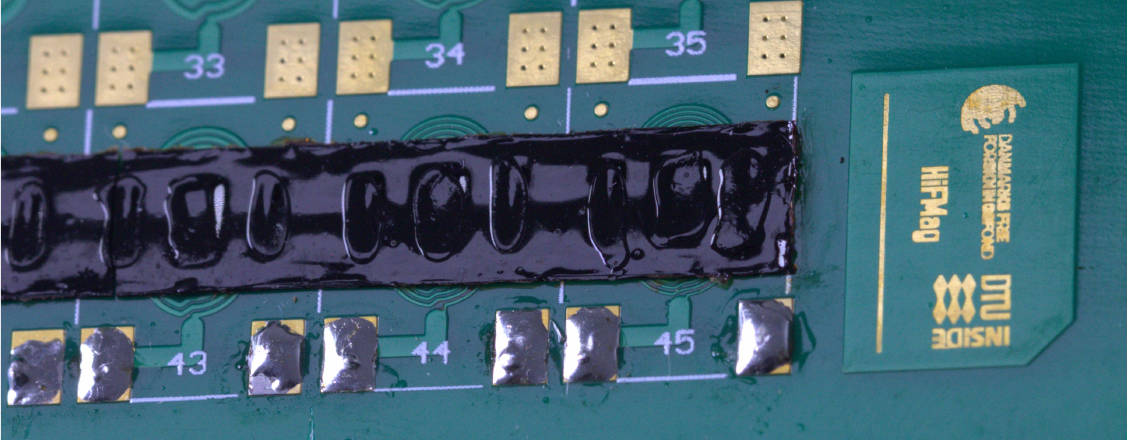}
    \caption{}
    \label{fig:a}
\end{subfigure}
\hfill
\begin{subfigure}{0.19\columnwidth}
    \centering
    \includegraphics[width=0.95\columnwidth,trim={8.5cm 10.5cm 8.5cm 10cm},clip]{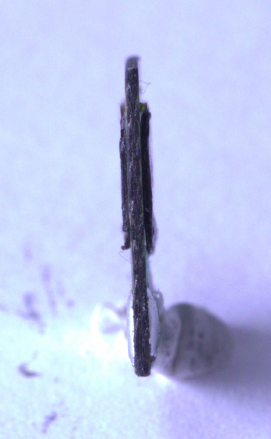}
    \caption{}
    \label{fig:b}
\end{subfigure}
\caption{Frontal (a) and side (b) view of the finished Print Circuit Board (PCB) based inductors. Core material is seen as dark brown, glossy polymer material on the green PCB background. Inductor and core size can be seen in figure \ref{fig:Printing}.}
\label{fig:ab}
\end{figure}

The presented nanocomposite with 45 vol\% particles was deposited as the magnetic core on a PCB-based inductor by manual printing / droplet casting, as seen in \textbf{figure \ref{fig:OverviewComic}} and \textbf{figure \ref{fig:Printing}}. The Inductor is a PCB based 3 turn inductor with 3 vias for the magnetic core. First, the backsides of the magnetic vias were blocked by tape and filled by droplet casting. After drying of the nanocomposite in the magnetic vias, a 300 micron thick upper magnetic core part was printed by use of masking tape and a layer-by-layer approach, i.e. manual printing. When the upper core part was dried the masking tape on the backside was removed and the same printing procedure was done on the backside to yield a closed, 3 legged magnetic core surrounding the 3 turn coil. The printing process is illustrated in \textbf{figure \ref{fig:Printing}} and the finished inductor can be seen in \textbf{figure \ref{fig:a}} and \textbf{\ref{fig:b}}.

\begin{figure}[tp!]
    \centering
    \includegraphics[width=0.95\columnwidth]{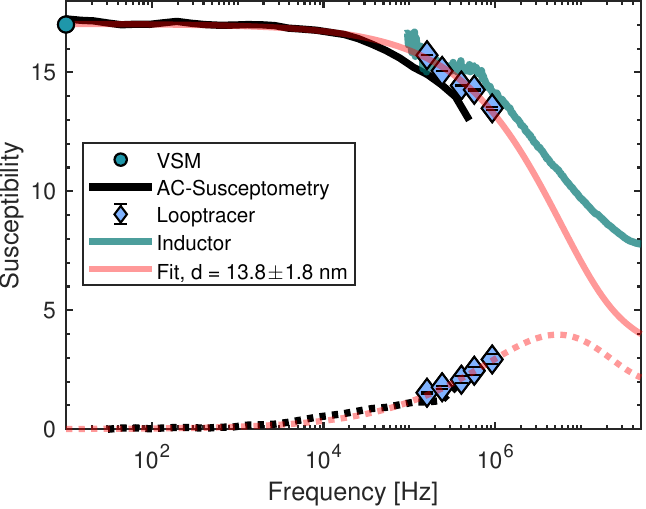}
    \caption{In- and out-of-phase component of the nanocomposite volume susceptibility from VSM, AC-susceptometry, looptracer-, and induction-measurements, together with Debye-model fit. Susceptibility was calculated from induction measurements by use of FEM-simulations and inductor core size seen in figure \ref{fig:Printing}. Debye-model fit resulted in particle diameter of 13.8$\pm$1.8 nm, anisotropy constant of 12.7 kJ/m$^3$, and superparamagnetic attempt time of 2 ns.}
    \label{fig:SusceptibilityOverview}
\end{figure}

The resulting inductor was successfully tested in a power converter operating at switching frequencies up to 3.5 MHz \cite{Bima_Preprint}.
Small signal inductance was measured by impedance analyser on the fabricated 3 turn inductor for frequencies between 100 kHz and 100 MHz, allowing to calculate susceptibility of the nanocomposite by use of FEM simulation and core-dimensions (see \textbf{figure \ref{fig:Printing}} and \textbf{figure \ref{fig:SusceptibilityOverview}}). A more detailed report on the design of inductor and converter for use of nanocomposites containing superparamagnetic maghemite particles as inductor core will be presented elsewhere \cite{Bima_Preprint}.

Susceptibility of the nanocomposite material found from VSM-, AC-susceptometry-, looptracer- and induction-measurements are shown in \textbf{figure \ref{fig:SusceptibilityOverview}} together with a fit to the Debye-model \cite{zambach2023b}.
It is seen that susceptibility above 850 kHz, found from inductions measurements, agrees with AC-susceptibility data and susceptibility found from hysteresis loop measured on the looptracer setup. The susceptibility above 1 MHz decreases with increasing frequency, seemingly levelling off in the high MHz range. The decrease is due to transition from SPM regime to blocked particles, the levelling off could be due to the frequency independent susceptibility of blocked single-domain nanoparticles \cite{zambach2023b}, or due to resonant frequency from the inductor. 
Discrepancies between the looptracer measurements and the inductor measurement are likely due to the different applied field used; the lowest excitation field used in the looptracer is several times higher than the field in the inductor when analysed by the impedance analyser. Another possible reason for discrepancies could be that the inductor measurement was performed in a closed flux configuration, i.e. no demagnetisation and that the demagnetisation constant in the AC-susceptibility and looptracer measurement could have changed due to changes in the sample mounting between the two measurements. Heat up during looptracer measurements also contributes, as higher temperature in the looptracer measurement could give lower susceptibility. We find that theoretic values for susceptibility, fits reasonably with the measured susceptibility for frequency below 4 MHz. Larger apparent volume fraction and saturation magnetisation than expected (360 vs 303 kA/m and 77 vs 44.7 vol\%) were found in the Debye model fit, which we attribute to dipolar interaction \cite{elfimova2019a}.

\section{Fabricational opportunities}
The here presented material offers increased flexibility with respect to design and fabrication of inductors, and both drop-casting and printing is possible. While nanocomposites containing nanoparticles in polymers like epoxy, PDMS, polysterene, and surfactants like oleic acid have been presented \cite{garnero2019a,shi2020a,hodaei2018a,yang2018a,kin2016a,kura2012a,kura2014a,yatsugi2019a,yun2014a,yun2016a}, the here presented material has fabricational advantages.
The use of maghemite in aqueous solution allow for use of PVA polymer, particle repulsion by surface charging, and independence of controlled atmosphere or glovebox, allowing easier large scale production \cite{garnero2019a,kin2016a}, and no need for annealing \cite{yatsugi2019a}. Most other presented fabrication methods for nanocomposites rely on compression molding to define inductor core structures \cite{yang2018a,li2022a,kin2016a,kura2012a,yun2014a}. We show that the here presented material can be printed directly on the inductor, making inductor fabrication and design much more flexible \cite{sanusi2022a,Bima_Preprint}. There have been examples of 3D-printable composite cores, however, these materials struggle with low particle loading, particle aggregation, and have been presented for particles larger than 50 nm only \cite{shi2020a,hodaei2018a}. The fabrication protocol presented here with pH-controlled repulsion between nanoparticles (to avoid the iso-electric point and associated aggregation) may be a stepping stone towards better filaments for 3D printing of soft magnetics. On a related note: it has been proposed that the most promising/feasible methods for larger scale inductor core production is to either coat the nanoparticles with silicon-oxide \cite{rowe2015a,li2022a,FeNi3_Article_Lu}, or to directly cast the particles in polymer/surfactant \cite{kura2012a,yun2016a}. Silicon oxide coating has, however, been proven to be troublesome, as usually larger, coated aggregates are formed rather than individual, coated particles \cite{FeNi3_Article_Lu}, and relative low particle loading is obtained \cite{rowe2015a,li2022a}. Particles in surfactant/oleic acid have shown promising, but does not result in a stable material, but a viscous liquid \cite{yun2016a}. A drawback of the presented material is shrinkage during drying / curing, which might result in flaking and brittle material for high particle loading, but slower curing/drying together with UV induced polymer cross-linking give satisfying material stability. Therefore, we believe the presented method demonstrate a way forward that has potential for large-scale fabrication. 

The effect of aggregation in nanocomposites for inductor core application has not been investigated thoroughly in the literature \cite{FeNi3_Article_Lu,yun2014a,yun2016a,yatsugi2019a,liu2005a,kura2012a,kura2014a,yang2018a,garnero2019a,shi2020a,hodaei2018a,yang2018a,li2022a,kin2016a,FeNi3_Article_Lu}.
In most of these works, the attempted procedure to achieve high susceptibility is by packing the magnetic nanoparticles as densely as possible, often using pressure to form a stable material \cite{yun2014a,yatsugi2019a,yang2018a,watt2018a}. This approach limits the design of later inductors and is more complicated compared to the approach presented here. Moreover, it may not produce the most optimal material properties \cite{zambach2023b}.
Only a few consider particle aggregation \cite{rowe2015a,garnero2019a}, but not in a quantitative manner, even though particle aggregation can increase superparamagnetic relaxation time and therefore in- and out-of-phase susceptibility of magnetic nanoparticles \cite{zambach2023b,Durhuus2025a}. 

There have been reports of low coercivity in DC magnetisation curves for nanocomposites \cite{garnero2019a,yun2016a,kura2012a,rowe2015a}, few studies measure both DC and AC susceptibility up to high kHz/low MHz range \cite{yun2016a}. It seems necessary, however, to measure both and supplement with AC-magnetometry at large fields (at least up to the field amplitudes relevant for applications), as there have been cases of composite materials that show soft magnetic behaviour for smaller applied fields, but show considerable hysteresis for larger applied fields \cite{FeNi3_Article_Lu,yatsugi2019a,liu2005a,yang2018a}.
An explanation could be that blocked single-domain particles appear magnetically soft for small applied fields, but show extensive hysteresis and high coercivity for larger applied fields \cite{komogortsev2021a}. Therefore, quantification of magnetic losses in the MHz range and at sufficiently large applied fields is necessary to prove that smaller, superparamagnetic particles can, in fact, be used for inductor core materials. 
New measurement techniques using on inductive compensation on two-winding inductors and AC-magnetometry, as done here, enable measurement of losses in early MHz range with $B$-field amplitude in the same range as micro-inductor core materials \cite{sanusi2022a,Bima_Preprint,TVeile2025}. This opens the possibility of fair comparison of susceptibility and losses in nanocomposite materials compared to conventional magnetic core materials, not done to date \cite{Bima_Preprint,TVeile2025}.

The material presented here has higher susceptibility than other low-coercivity nanocomposites using maghemite or other iron oxide particles \cite{yun2014a,rowe2015a,watt2018a}, and slightly lower susceptibility than values reported for nanocomposites using zinc-ferrite nanoparticles \cite{yun2016a}. Materials using nanometer-sized, metallic particles have not yet been able to reach higher susceptibility, as they often have high degree of particle aggregation and very large coercivity \cite{yatsugi2019a,yang2018a}. It should, however, be possible to obtain high susceptibility in nanocomposites containing FeNi, Fe, FeCo particles, if well-dispersed \cite{zambach2023b}. In order to be competitive with ferrites as inductor core material, also a saturation magnetisation above 398 kA/m ($\mu_0 M>$ 500 mT) is desirable \cite{he2023a}. The here presented material has the saturation magnetisation of the volume fraction of the particles times 303 k/Am ($\mu_0 M$ = 381 mT). Thus we predict that in order to be competitive with ferrites nanocomposites using SPM particles of higher moment materials, like FeNi, Fe or Co, need to be investigated. 

Considering the use of larger moment particles it is important to investigate the effect of dipolar interactions on the superparamagnetic relaxation as AC-behaviour is strongly influenced by particle distance for larger, metallic particles \cite{liu2005a,FeNi3_Article_Lu}. For particles with lower magnetic saturation materials, like maghemite used here, there is only little effect of dipolar interaction on the SPM relaxation time, even at high volume fractions of particles \cite{yun2014a,yun2016a}.

\section{Conclusions}
We have demonstrated that it is possible to synthesise a printable and castable nanocomposite containing well-dispersed superparamagnetic maghemite nanoparticles of diameter 11$\pm$3 nm with volume fractions of 10-45 percent in a polyvinyl alcohol matrix. The material shows no hysteresis at room temperature when measured by vibrating sample magnetometry, and reaches a sample volume susceptibility of 17 for 45 vol\% of particles with a saturation magnetisation of around 135 kA/m ($\mu_0 M = 170$ mT). AC-susceptometry shows that the in-phase susceptibility is flat up to 10 kHz, above which a decrease/increase in in-phase/out-of-phase component occurs due to the polydispersity of the sample (i.e. the fraction of larger particle sizes). This can be counteracted by use of narrower particle size distribution. We have measured high-frequency hysteresis loops from 160.6 to 922.7 kHz, with induced $B$-fields in the sample ranging  7-107 mT, which shows that power loss in the material is on the order of $10^3$-$10^4$ kW/m$^3$ for 30 mT $B$-field. Fitting based on a Debye model yields particle diameter distribution of 13.8$\pm$1.8 nm and predict low field power losses well. The power loss depends on $B$-field squared, whereas for frequency a smaller exponent around 1.0-1.3 is found. This shows that in comparison to bulk materials, eddy-currents are minimised and only hysteresis from larger particles transitioning to blocked state is present in the material, even at high $B$-field. Narrower particle diameter distribution could result in considerably lower losses. The susceptibility of the nanocomposite could potentially be improved by aligning the particle anisotropy axes with respect to the applied field or by tuning inter-particle interactions \cite{zambach2023b}. 

The presented material is easily deposited on PCB-inductors by drop-casting or manual printing, allowing for fast and flexible inductor fabrication, and it can be used directly after drying / UV-curing, without need for annealing, compression molding or use of protective atmosphere. The material can therefore be integrated directly in existing micro/nano-fabrication methods, as demonstrated by the fabrication of PCB based 3-turn inductors, on which flat, 3 legged, closed E-cores were deposited. 

Future material research should also investigate the use of monodisperse metallic superparamagnetic particles with higher saturation magnetisation than maghemite for inductor core applications, as higher saturation magnetisation give much larger particle susceptibility \cite{zambach2023b}.

\FloatBarrier
\section{Experimental Section}
\paragraph{\textbf{Nanoparticle synthesis and characterisation}}
\textit{Chemicals}: 
Iron (II) chloride tetrahydrate (FeCl$_2$·4H$_2$O, $\geq99.0$\%), iron (III) chloride hexahydrate (FeCl$_3$·6H$_2$O, 98.0-102\%), iron (III) nitrate nonahydrate (Fe(NO$_3$)$_3$·9H2O, $>99$\%), sodium hydroxide (NaOH, $\geq98$\%), diethylene glycol (DEG, 99\%), N-dymethyldiethanolamine (NMDEA, $\geq99$\%), nitric acid (HNO$_3$, 69\%), ethanol (99.8\%), acetone ($\geq99.8$\%) and ethyl acetate (99.5\%). All products were used without further purification.

\textit{Synthesis of $\gamma$-Fe$_2$O$_3$ nanoparticles}: 
The nanoparticles were synthesized using a polyol process following a previously established methodology \cite{hugounenq2012a,caruntu2004a}. A mixture of two solvents, diethylene glycol (DEG) and N-methyldiethanolamine (NMDA) (1:1 by weight), was used.  Typically, 4 mmol of FeCl$_3$·6H$_2$O and 2 mmol of FeCl$_2$·4H$_2$O were dissolved in 80 mL of polyol mixture, and the solution was stirred at room temperature for 1h. Separately, 16 mmol of NaOH were dissolved in 40 g of polyol mixture, and this solution was added to the solution of iron chlorides and stirred for another 3 h. The temperature of the solution was then elevated to 220 $^{\circ}$C using a ramp temperature of 2 $^{\circ}$C/min and stirred for 18 hours. After a slow cooling down by removing the heating system, the sediments were separated magnetically and washed 3 times with a mixture of ethanol and ethyl acetate (1:1 by volume) and once with 10\% nitric acid. The nanoparticles were further oxidized to maghemite by adding 20 mL of water solution containing 8.25 g of iron (III) nitrate, and heating the mixture to 80 $^{\circ}$C for 45 min. The nanoparticles were washed again, once with 10\% nitric acid and 3 times with acetone. Finally, the nanoparticles were dispersed in 6 mL of pure water. The solution was stable in acid or basic conditions for several months. The final nanoparticle solution used in the experiments was a mixture of 8 different batches prepared as described above. All batches were prepared using the same amounts of precursors and solvents. 

\textit{Transmission electron microscopy (TEM)}:
A FEI Tecnai G2 T20 transmission electron microscope operated at 200 kV was used in this study. TEM sample preparation was done by putting a droplet of the diluted nanoparticle suspension on a carbon film copper grid and allowing it to dry in air. Images were acquired in bright field (BF) imaging mode.

\textit{Inductively coupled plasma mass spectrometry (ICP-MS)}:
A Thermo Fisher Scientific iCAP Q was used to determine the iron content of the sample. Sample digestion: 90 $\mu$L of sample were digested in 1 ml HNO$_3$ Suprapur 65\%. This sample was diluted twice in HNO$_3$ Suprapur 2\%, first a 1000 times and after about 130 times.

Particle saturation magnetisation of 303$\pm$14 kA/m at 1 Tesla (62$\pm$3 Am$^2$/kg with a maghemite density of 4880 kg/m$^3$) was measured by Vibrating Sample Magnetometry as average of 5 measurements for each of 5 liquid samples, using the iron-concentration values found by ICP-MS.

\paragraph{\textbf{Nanocomposite synthesis and characterisation}}
Nanocomposites of different particle fraction were obtained by adding the concentrated nanoparticle solution (44.1 mg Fe/mL), PVA solution (16.7 wt\% 9000-10000 M$_w$ polyvinyl alcohol in H$_2$O), and photoinitiator (Darocur 1173) with the amounts given in \textbf{table \ref{tab:CompositeVol}}. 
\begin{table}[tp!]
\centering
 \caption{Volumes of nanoparticle solution, polyvinyl alcohol solution and photo-initiator used for nanocomposite synthesis.}
 \label{tab:CompositeVol}
 \begin{ruledtabular}
  \begin{tabular}{cccc}
    \textrm{Sample} & \textrm{Particle } & \textrm{PVA} & \textrm{Darcur}\\\textrm{[vol\%]} & \textrm{solution [ml]} & \textrm{solution [ml]} & \textrm{1173 [$\mu$l]}\\
    \colrule
    1 & 0.4  & 3 & 168 \\
    4.5 & 1.25  & 2 & 112 \\
    10.1 & 3  & 2.25  & 126 \\
    17.7 & 3.6  & 1.44  & 81 \\
    35.3 & 3  & 0.6  & 33.6 \\
    44.7 & 3.6  & 0.54 & 30.2 
  \end{tabular}  
\end{ruledtabular}
\end{table}

After mixing of the solutions and photo-initiator the samples were UV-cured under a Hg-lamp for 2 days, after which they were further dried at 50 degrees for 2 days. Droplet-cast / screen-printed samples were placed under UV-lamp for 2 hours, between each deposition, no further drying was required due to faster evaporation time.

\textit{Small angle neutron Scattering (SANS)}:
Neutron scattering experiments were performed with the Yellow Submarine instrument at the Budapest Neutron Centre (BNC) using neutron wavelengths of 4.2 and 9.7 Å and detector distances of 1.15 and 5.26 meters and 10 mm beam diameter \cite{alm2021a}. For further information and full model description see supporting information.

\textit{Vibrating sample magnetometry (VSM)}:
DC-magnetisation measurements were performed on a Lake Shore Cryotronics 7407-series VSM. Measurements were taken at room temperature with applied field up to 1 Tesla, using the point-by-point mode, step size of 795.8 A/m (10 Oe), 0.1 second hardware time constant and 2 seconds software averaging time.
Disk-samples were oriented with diameter parallel to the applied field. The sample demagnetisation factors are specified in \textbf{table \ref{tab:VSMDemag}}.
\begin{table}[tp!]
\centering
 \caption{Sample weight, size (disk diameter and height), and demagnetisation factor along the diameter N$_{\textrm{d}}$ for samples used for VSM measurements.}
 \label{tab:VSMDemag}
 \begin{ruledtabular}
  \begin{tabular}{cccc}
    \textrm{Sample [vol\%]} & Weight [mg] & \textrm{Size (d$\times$h) [mm]} & \textrm{N$_{\textrm{d}}$}\\
    \colrule
    1 & 217.8 & 11.8$\times$1.92 & 0.1055 \\
    4.5 & 218.7 & 11.8$\times$1.70 & 0.0955\\
    10.1 & 215.4 & 11.9$\times$1.49 & 0.0800\\
    17.7 & 13.5 & 6$\times$0.28 & 0.0350\\
    35.3 & 9 & 6.2$\times$0.13 & 0.0155\\
    44.7 & 9.8 & 4.7$\times$0.20 & 0.0325
  \end{tabular}  
\end{ruledtabular}
\end{table}

\textit{AC-susceptibility}:
Susceptibility measurements between 10 Hz and 500 kHz were performed on a DynoMag AC-susceptometry system from RISE Research Institutes of Sweden, using 0.5 mT applied field. Disk samples were oriented parallel with applied field. Demagnetisation factors used in the AC-susceptibility and looptracer measurements can be seen in \textbf{table \ref{tab:AC_Demag}}.
\begin{table}[tp!]
\centering
 \caption{Sample weight, size (disk diameter and height), and demagnetisation factor along the diameter N$_{\textrm{d}}$ for samples used for AC-susceptometry and looptracer measurements.}
 \label{tab:AC_Demag}
 \begin{ruledtabular}
  \begin{tabular}{cccc}
    \textrm{Sample [vol\%]} & Weight [mg] & \textrm{Size (d$\times$h) [mm]} & \textrm{N$_{\textrm{d}}$}\\
    \colrule
    1       & 42.9     & 6$\times$1.50      & 0.1455\\
    4.5     & 57.2     & 6$\times$1.70      & 0.1640\\
    10.1    & 26.4     & 6$\times$0.67      & 0.0770\\
    17.7    & 13.5      & 6$\times$0.28     & 0.0350\\
    35.3    & 9         & 6.2$\times$0.13   & 0.0155\\
    44.7    & 9.8       & 4.7$\times$0.20   & 0.0325
  \end{tabular}  
\end{ruledtabular}
\end{table}

\textit{Looptracer}:
High-field dynamic magnetisation measurements were done at 160.6, 241.0, 404.0, 570.6, and 922.7 kHz on a self-built system using a magneTherm system from nanoTherics Ltd as excitation source and two counter-would pick-up coils, and Zürich Instruments MFLI 5 MHz lock-in amplifier \cite{TVeile2025}. Applied field amplitudes of 0.8, 4, 8, and 12 kA/m were used. The system was calibrated to moment and phase by using a small secondary coil excited by the lock-in amplifier tone generator, and by use of paramagnetic Dy$_2$O$_3$ powder sample for each frequency and applied field.

\textit{Inductance measurements}: 
Inductance measurements were performed on a Agilent/HP 4294A Precision Impedance Analyzer, using a oscillator level of 500 mV. The analyser was calibrated by a shorted PCB having the same dimensions as the inductor, but without the 3 turn coil. Inductance was measured with and without magnetic core in the range of 1 kHz to 100 MHz. Susceptibility in the range 0.8-50 MHz was calculated with help of FEM simulations described below.

\textit{FEM simulations}: Software used for FEM simulations was COMSOL Multiphysics 6.1 with magnetic field solver. Sizes of the inductor and core were set as seen in figure \ref{fig:Printing}, mesh 1/5 of relevant skin depth at 10 MHz. Sweeping core material permeability $\mu_\textrm{r} = (1+\chi)$ from 1 to 100 the resulting inductances (mutual- and self-inductance) were used to calculate increase of inductance with respect to the case of no inductor core. This was then used with inductance measurements on the 3 turn inductor (with and without magnetic core) to find the permeability (susceptibility) of the deposited magnetic core material.

\section*{Acknowledgments}
The authors thank the Independent Research Fund Denmark (project HiFMag, grant number 9041-00231A), the Danish Agency for Science, Technology, and Innovation (instrument center DanScatt, grant number 7129-00006B) and the Taumose Legat (grant 2022) for financial support.

\FloatBarrier
\bibliography{Bibliography}

\clearpage

\hspace{-2cm}
\begin{minipage}{21cm}
    \centering
    \includegraphics[width=21cm,page=1]{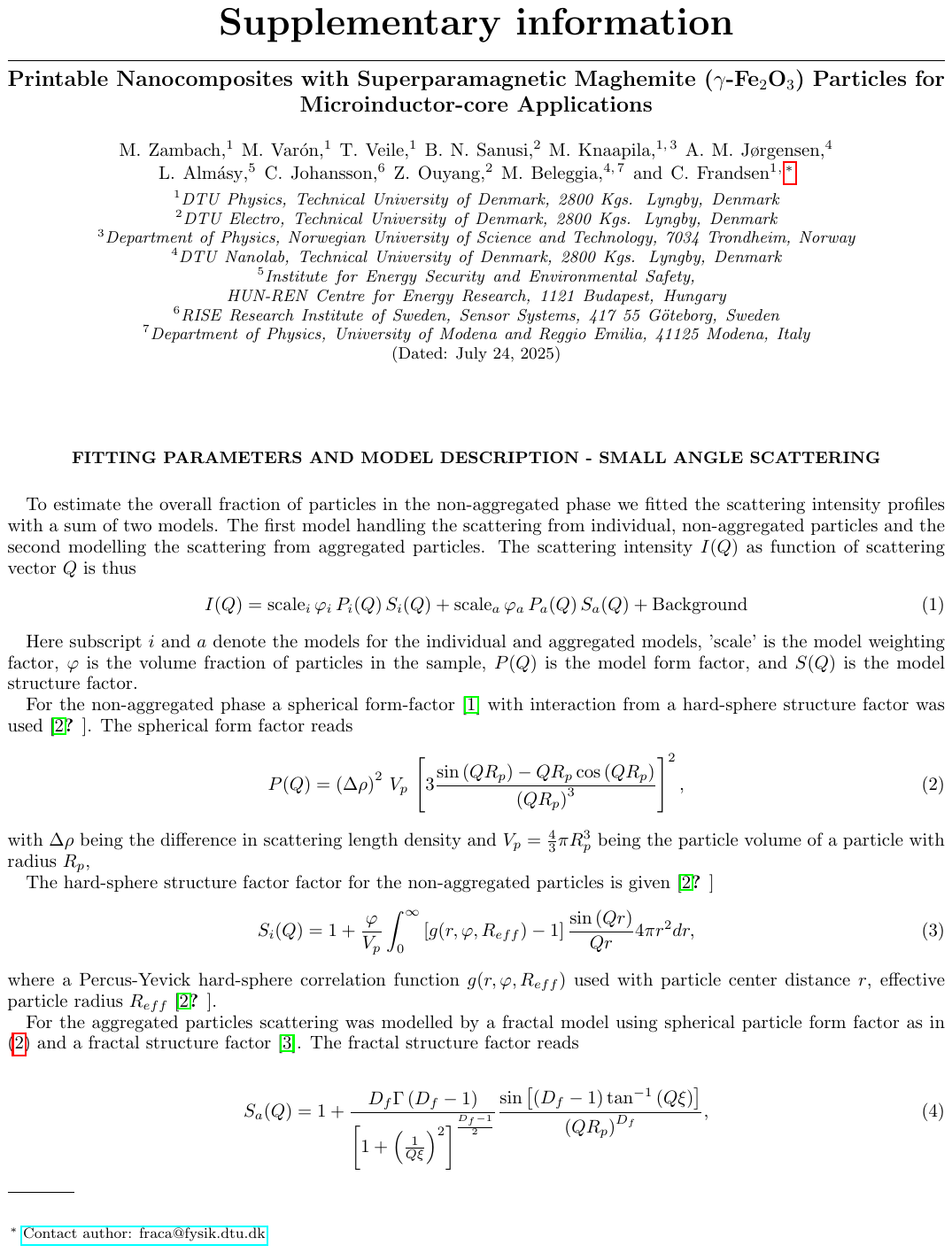}
\end{minipage}
\hfill

\clearpage

\hspace{-2cm}
\begin{minipage}{21cm}
    \centering
    \includegraphics[width=21cm,page=2]{Article_2___Material_Paper__Supplementary___CleanVersion_.pdf}
\end{minipage}
\hfill

\clearpage

\hspace{-2cm}
\begin{minipage}{21cm}
    \centering
    \includegraphics[width=21cm,page=3]{Article_2___Material_Paper__Supplementary___CleanVersion_.pdf}
\end{minipage}
\hfill

\end{document}